\documentclass[preprint,12pt]{elsarticle}
\usepackage{amssymb}
\usepackage{amsmath}

\usepackage{siunitx}
\usepackage{mathtools}
\DeclareSIUnit\sample{S}
\newcommand{\Dphi}{\Delta\phi}
\newcommand{\of}{\left(f\right)}
\newcommand{\ot}{\left(t\right)}
\newcommand{\oeps}{\left(\varepsilon\right)}
\newcommand{\res}[1][]{_{\text{res}\if\relax\detokenize{#1}\relax\else,{#1}\fi}}
\newcommand{\Df}{\Delta f}

\newcommand{\tx}{_\text{tx}}
\newcommand{\rx}{_\text{rx}}
\newcommand{\ex}{_\text{exp}}

\DeclareMathOperator{\argtanh}{argtanh}
\DeclareMathOperator{\sign}{sign}
\DeclareMathOperator{\Ftf}{\mathcal{F}}

\usepackage{hyperref}
\hypersetup{colorlinks=true,linkcolor=blue,anchorcolor=red,citecolor=blue,filecolor=red,urlcolor=red,pdfauthor=Jan Kober}

\makeatletter
\def\ps@pprintTitle{%
 \let\@oddhead\@empty
 \let\@evenhead\@empty
 \def\@oddfoot{\footnotesize
      \parbox{\textwidth}{\centering \itshape
       © 2026. This manuscript version is made available under the CC-BY-NC-ND 4.0 license \\
       \url{https://creativecommons.org/licenses/by-nc-nd/4.0/}}}%
 \let\@evenfoot\@oddfoot}
\makeatother

\begin{document}

\begin{frontmatter}

\author[itcas]{Jan Kober}
\author[itcas,fnspe]{Radovan Zeman\corref{corr}}
\ead{rzeman@it.cas.cz}
\author[polito]{Marco Scalerandi}
\cortext[corr]{Corresponding author.}

\affiliation[itcas]{Institute of Thermomechanics, Czech Academy of Sciences, Prague, Czechia}
\affiliation[fnspe]{Faculty of Nuclear Sciences and Physical Engineering, Czech Technical University in Prague, Czechia}
\affiliation[polito]{DISAT, Condensed Matter Physics and Complex Systems Institute, Politecnico di Torino, Italy}

\title{Fixed-phase Resonance Tracking for Fast Nonlinear Resonant Ultrasound Spectroscopy} 

\begin{abstract}
Nonlinear Resonant Ultrasound Spectroscopy (NRUS) experiments that rely on repeated sampling of resonance curves are inherently sensitive to measurement protocol due to evolution of material parameters caused by fast and slow dynamic effects. We introduce a model-assisted discrete-time resonance tracking method that maintains a system at its instantaneous resonance condition without the need to acquire full frequency sweeps. Resonance is defined through a prescribed phase relation between excitation and response, and the excitation frequency is iteratively updated using a linearized frequency--phase model. The procedure allows controlled suppression of transient wave buildup using optional feedforward correction with respect to an external control parameter. The method is demonstrated on NRUS and performing a conditioning--relaxation protocol, both conducted on a sandstone bar, providing estimates of resonance frequency and damping. Comparison with conventional approaches shows that measurement speed and mode stability significantly influence the inferred nonlinear indicators. The proposed framework is not limited to nonlinear acoustics and can be applied to arbitrary resonant systems with slowly evolving parameters.
\end{abstract}


\end{frontmatter}

\section{Introduction}\label{sec:introduction}

Nonlinear elasticity is observed  in a broad range of materials (rocks \cite{johnson_nonlinear_1996,riviere_frequency_2016,remillieux_decoupling_2016,li_nonlinear_2018}, concrete and mortar \cite{bentahar_hysteretic_2006,payan_quantitative_2014,dominguez-bureos_stress-_2025}, glass beads \cite{yoritomo_slow_2020-2},  metal alloys \cite{kober_elastic_2021,kamali_influence_2019}, etc.). With respect to the variability of the materials microstructure, different physical sources and types of nonlinearity are always present. Classical nonlinearity \cite{landau_theory_1986}, hysteresis \cite{guyer_hysteresis_1999} and slow dynamics \cite{tencate_slow_1996,scalerandi_conditioning_2018} have been often observed at the same time with different proportions depending on a particular material state, e.g., it has been shown that dislocations in the crystal lattice are responsible for classical and hysteretic response \cite{barsoum_dynamic_2005} and that fast and slow dynamics can lead to effects associated with both hysteresis and classical nonlinearity \cite{zeman_distribution_2025,remillieux_decoupling_2016}. By the assessment of material's elastic nonlinearity, information can be gained on the microstructural properties of the material (e.g. grain size \cite{kober_material_2022} and dislocation density \cite{nagy_fatigue_1998}) and developing damage (cracks \cite{ulrich_interaction_2007,jin_dynamic_2018}, embrittlement \cite{kober_monitoring_2026}, heat damage \cite{payan_applying_2007}, delamination \cite{segers_backside_2020}, etc.).

The development of experimental approaches that are reliable and accurate in the lab and extendable to field applications is of importance for both experimental and theoretical reasons. Various techniques have been proposed: harmonics evaluation \cite{van_den_abeele_nonlinear_2000-1}, coda wave interferometry \cite{hadziioannou_stability_2009}, Scaling Subtraction Method \cite{bruno_analysis_2009,ohara_enhancement_2012}, wave mixing \cite{kober_assessing_2020,croxford_use_2009}. Among others, Nonlinear Resonance Ultrasound Spectroscopy (NRUS) \cite{van_den_abeele_nonlinear_2000,remillieux_resonant_2015,lott_local_2016,geimer_nonlinear_2023,gregg_electromagnetic_2020} has several advantages, because the procedure is quantitative, easy to implement and very sensitive to microstructural changes in the material. The typical measurement procedure is based on measuring the resonance curve around one of the longitudinal modes by sweeping the frequency at a given amplitude of excitation and increasing the drive amplitude to quantify the dependence of resonance frequency and damping on strain amplitude. Note that NRUS can be used also for shear, flexural or torsional modes \cite{kim_flexural_2025} and it can be easily applied for long time monitoring \cite{kober_monitoring_2026}.  NRUS provides averaged information over the entire specimen volume, thus it is not particularly suitable for localization, even though a few modifications of the procedure were introduced to allow it \cite{liu_localization_2010,bentahar_nonlinear_2020}.  

The procedure requires a long acquisition time, taking seconds per frequency sweep (given by the time required to reach stationary state), thus the temporal resolution is limited and the material evolution during the frequency sweep and the whole experiment creates a strong bias which depends on the duration \cite{kober_role_2025}. Also the sample is probed with different heterogeneity levels (due to transition among different strain profiles when sweeping frequency). Probing the material with monochromatic waves at fixed frequency and increasing amplitude \cite{mechri_separation_2019} is faster, but the method does not strictly test the "same" material at each amplitude (again due to different strain profiles).

The goal here is to propose an alternative approach consistent with the conventional NRUS, which significantly reduces measurement duration and assures consistent strain distribution at each driving amplitude. These qualities are particularly important in the assessment of materials exhibiting significant slow dynamic response. Also applications, where serial sample testing or condition monitoring are performed, will benefit from adopting the method.

In case of  slow dynamic response, the material parameters do not remain constant during the acquisition time in any experimental protocol. In a typical NRUS experiment, the material modulus and damping evolve on time scales comparable to, or even longer than, the duration of a single frequency sweep, leading to resonance curves that do not correspond to a stationary system \cite{kober_role_2025}. Slow dynamics in materials like sandstone spans a remarkably wide spectrum of relaxation times, ranging from microseconds \cite{zeman_distribution_2025} up to hours or days \cite{lebedev_experimental_2024}, thus the effects arise within a period of any probing wave and easily outlast the duration of any standard acoustic experiment. 
Two factors contribute to the complexity of the assessment. First, it is spatial heterogeneity as the strain-induced modulus and damping variations depend on the position along the sample and persist even after the excitation ends. Consequently, probing at different frequencies means probing effectively different materials. Second, the temporal evolution contributes to the bias since the local material state at any point depends on the full history of strain amplitude and duration it has experienced. 

Consequently, the measured material parameters depend not only on the excitation amplitude but also on the experimental protocol. In particular, in resonance experiments the estimated resonance frequency and attenuation are sensitive to the sweep rate or the frequency step size, the order in which the amplitudes and frequencies are applied and the time allowed for reaching standing wave condition at each frequency. When probing a hysteretic nonlinear elastic material to determine its elastic properties (and/or their time dependence), it might be not well defined which material is effectively tested. This makes it difficult to understand the physics responsible, going beyond phenomenological models currently developed and based on multirelaxation theory \cite{sens-schonfelder_model_2018,zeman_distribution_2025,yoritomo_slow_2025} or other approaches \cite{bittner_mechanistic_2021,jin_integrated_2020,li_nonlinear_2018,lebedev_slow_2023}. At the same time, the quantification of nonlinearity might be biased, which is of interest because of practical application in the field of materials \cite{kim_experimental_2006,kersemans_nondestructive_2014,remillieux_resonant_2015} and buildings \cite{gueguen_nonlinear_2016} characterization. Nonlinear elastic parameters are indeed very sensitive to the presence of microstructural changes \cite{jin_dynamic_2018,scalerandi_nonlinear_2013,kim_measuring_2017,gao_effect_2023,chavazas_impact_2024,feng_effects_2022,choi_comparison_2019,kober_material_2022,renaud_anisotropy_2013,zeman_relaxation_2024}. 

Unlike in the conventional approach, we aim to probe the sample using, for each drive amplitude, a single continuous excitation. The excitation frequency is selected, so that it tracks the resonance evolution due to the change in amplitude and slow dynamics. This reduces the measurement time from typically seconds per frequency sweep to only the time required to reach standing wave conditions. For this purpose, a resonance tracking procedure exploiting the phase properties of the resonance curve is developed involving an iterative correction of the deviation from resonance and prediction for the following amplitude, with an approach conceptually similar to others often used in nonlinear dynamics. 

Phase-based frequency tracking has indeed been employed for  following the resonance condition as the oscillation amplitude increases. The resulting backbone curve characterizes the intrinsic nonlinear behavior of the system \cite{vakakis_normal_2001, kerschen_tracking_2016,nayfeh_nonlinear_1995}. Considering that resonance corresponds to a specific phase relation between excitation and response, these approaches have been widely used in nonlinear modal analysis, micro- and nano-mechanical resonators, and atomic force microscopy \cite{rhoads_nonlinear_2008,antonio_frequency_2012}. Phase-locked loop (PLL) techniques are frequently employed to track the backbone curve by continuously adjusting the excitation frequency to maintain a prescribed phase condition \cite{peter_excitation_2017,denis_identification_2018}. 

PLL-based approaches provide efficient frequency synchronization; however, they are primarily control-oriented. In contrast, the method proposed here is formulated as a discrete-time, model-assisted resonance tracking procedure tailored to resonant spectroscopy. The excitation frequency is updated iteratively based on the measured phase condition, while explicitly accounting for transient wave buildup and incorporating optional feedforward terms associated with controlled variations of experimental parameters. The objective is not merely frequency locking, but controlled tracking of a resonance condition suitable for extracting material parameters even in the presence of evolving elastic properties.

The method to track the resonance frequency at varying amplitude is discussed in Section \ref{sec:method}. In Section \ref{sec:demo}, experimental details are given together with some experimental results. In Section \ref{sec:applications} some applications are discussed, in particular for what concerns NRUS, conditioning and relaxation monitoring. Conclusions and discussion will finally be given. 

\section{Method}\label{sec:method}

\subsection{Nonlinear Resonant Ultrasound Spectroscopy}\label{sec:nrus}

\subsubsection{Amplitude-based resonance frequency estimation}

\begin{figure*}
	\centering
		\includegraphics[width=\textwidth]{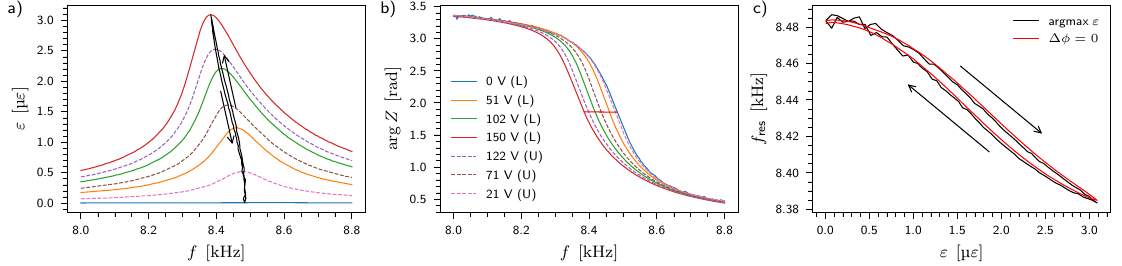}
	\caption{Experimental data from NRUS, selection of excitation amplitudes from loading (L) and unloading (U), a) strain amplitude vs. frequency, b) phase vs. frequency. Comparison of resonance frequency obtained fitting the maximum of the amplitudes (black) and that obtained using the phase-based definition (red).}
	\label{fig:nrus}
\end{figure*}

Nonlinear Resonant Ultrasound Spectroscopy (NRUS) is commonly implemented by measuring the frequency response of a selected resonance mode from which the resonance frequency (position of the maximum) and attenuation (often derived from the peak width) are extracted. By repeating such measurements, sweeping over frequency (around a chosen mode) at different excitation amplitudes, amplitude-dependent shifts of the resonance frequency and changes in attenuation are used as indicators of elastic nonlinearity. Often the excitation amplitude protocol includes a loading phase (amplitude increases) followed by an unloading phase (amplitude decreases).

The typical results of an NRUS measurement are illustrated in Fig. \ref{fig:nrus}, obtained on a sandstone sample excited at its first longitudinal mode, where both the loading and unloading branches are measured. The difference in loading and unloading amplitude dependences and the shapes of the peaks themselves indicate slow dynamic behavior. As a consequence, the measured material parameters depend not only on the excitation amplitude, but to some extent on the experimental protocol as well.

The use of chirp in NRUS, rather than sweeping over frequency with monochromatic waves allows to shorten measurement duration, but does never ensure stationary standing wave conditions, being the sample only consistently close to them during wave propagation \cite{maier_noncontact_2018}. Provided the signal is long enough, transient waves contribute only slightly and chirp signals allow to probe the sample equivalently to a sweep over frequency using monochromatic waves for what concerns the resonance frequency, while some issues might arise for the quantification of the Q factor. 

Despite the limitations, the conventional approach to NRUS using frequency sweeps remains widely used because it provides a direct visualization of the resonance behavior and allows a straightforward extraction of amplitude-dependent indicators. However, the protocol dependence discussed above raises the question of whether measuring the full resonance curve is necessary (or even whether it is correct) to define the resonance frequency itself, or whether equivalent information can be obtained in a way that is less time-consuming and thus leads to reduction of the protocol-dependent slow dynamic effects. 

\subsubsection{Phase-based resonance frequency estimation}\label{sec:phase-based}

An alternative theoretical definition of resonance is the phase difference between the excitation and the measured response (Fig. \ref{fig:nrus}b). While the shape of the resonance curves is affected by elastic nonlinearity and slow dynamics, the phase exhibits a well-defined and reproducible crossing of a reference value at the resonance. The experimental data demonstrate that the frequency at which the phase crosses this reference value coincides with the frequency of the maximum response amplitude (Fig. \ref{fig:nrus}c) and can be used as a robust means of identification of the resonance frequency (compare the noise in the estimation based on the amplitude and phase in Fig. \ref{fig:nrus}c). Importantly, this definition remains meaningful even when the resonance curve itself is distorted or evolving due to nonlinear and memory effects (slow dynamics). The specific value of the phase at resonance deviates from its theoretical value as it depends on which physical quantities are measured and on the measurement configuration; a detailed description of how this reference phase is determined is provided in Section \ref{sec:calibration}. This phase-based definition of resonance forms the basis for the resonance tracking procedure in Section \ref{sec:rt}.

\subsection{Analytical model of resonance response}\label{sec:MoDaNE}

It is possible to express the steady state responses of amplitude and phase dependences on frequency analytically for a linear elastic material. Here we follow the Modulus and Damping Nonlinearities Evaluation (MoDaNE) \cite{mechri_separation_2019,di_bella_analysis_2019} approach.

We adopt a normalized formulation of the MoDaNE equations. Rather than working with absolute displacement amplitudes (requiring scaling of amplitude-related parameters), we use the complex transfer function between transmitted and received signals ($y\tx$ and $y\rx$). This normalization eliminates the dependence of parameters on the drive amplitude, allowing the intrinsic resonance frequency and damping to be estimated solely from the measured phase and normalized amplitude. More specifically, from the Fourier components of the source and response signals calculated at the excitation frequency $f$ using DFT, we define
\begin{equation}
    Z\of \equiv \frac{\Ftf\left[y\rx\right]\of}{\Ftf\left[y\tx\right]\of} = a\of\exp j\varphi\of.\label{eq:Z}
\end{equation}

\begin{figure}
	\centering
		\includegraphics[width=0.7\columnwidth]{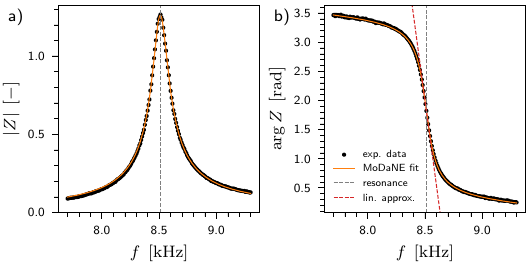}
	\caption{Resonance curve (experimental data) fitted using MoDaNE (calibration included), a) amplitude and b) phase vs. frequency. A linear approximation of the phase curve at resonance is shown as a red dashed line.}
	\label{fig:res}
\end{figure}

When a bar of length $L$ is excited at one end ($x=0$) by a monochromatic forcing of frequency $f$ (with zero initial phase) near its $n$th longitudinal mode, assuming free--free boundary conditions, the normalized displacement of the other end ($x=L$) is $u\ot=a\exp j\left(2\pi f t+\varphi\right)$ with the amplitude and phase given by
\begin{align}
    a\of&=\frac{U_0}{\sqrt{\cosh^2\left(\alpha L\right)-\cos^2 \left(\pi f/f\res\right)}}, \label{eq:MoDaNE_a_theory}\\
    \varphi\of&=\pi n-\arctan\left[\frac{\tan\left(\pi f/f\res\right)}{\tanh \left(\alpha L\right)}\right],\label{eq:MoDaNE_phase_theory}
\end{align}
\noindent where $\alpha$ is a damping coefficient and $f\res$ is the resonance frequency, $U_0$ is the normalized amplitude at $x=0$. The solution predicts that the displacements of the bar ends at resonance oscillate with either same or opposite phase depending on the mode number. The agreement of the analytical solution with experimental data is shown in Fig. \ref{fig:res}. Note that the fitting includes calibration terms accounting for frequency dependent amplitude and phase response of the measurement chain (see Section   \ref{sec:calibration}  for details).

Inversion of Eqs. \ref{eq:MoDaNE_a_theory} and \ref{eq:MoDaNE_phase_theory} enables estimation of resonance frequency and damping based on a measurement of phase $\arg Z$ and amplitude $\lvert Z\rvert$ of a single monochromatic excitation at a given frequency as
\begin{align}
    f\res&=\frac{f}{1-\frac{\sign\sin\left(\varphi-\pi n\right)}{\pi n}\arctan{\sqrt{r/2\cos^2\varphi}}},
    \label{eq:mod_inv_a} \\
    \alpha&=\frac{1}{L}\argtanh{\sqrt{r/2\sin^2\varphi}},
    \label{eq:mod_inv_p}
\end{align}
\noindent where
\begin{equation}
    r=-\frac{a^2}{U_0^2}-\cos 2\varphi+\sqrt{1+\frac{a^4}{U_0^4}+2\frac{a^2}{U_0^2}\cos 2\varphi}.
\end{equation}
We emphasize that MoDaNE describes a stationary standing wave, thus the transient oscillation buildup following a change in excitation amplitude or frequency must subside before data acquisition.

It is important to note that while the MoDaNE framework is derived for linear elastic media, its validity is not a prerequisite for the execution of the proposed method in the nonlinear regime. The model serves two distinct purposes. First, it provides a baseline calibration of the experimental setup at low strain amplitudes where linear behavior holds. Second, during nonlinear tracking, the equations enhance estimates and boost convergence based on measurement already performed almost at resonance. This is further discussed in Section \ref{sec:discussion-modane}.

\subsection{Resonance tracking}\label{sec:rt}

\subsubsection{Phase feedback and amplitude feedforward tracking}

Let us consider a sample excited with a monochromatic wave at driving frequency $f$. As long as the excitation frequency $f$ is matched to the resonance frequency $f\res$, the observed phase is equal to the resonance value. When the material parameters evolve, the resonance frequency shifts and a non-zero phase difference emerges, denoted as $\Dphi\equiv\arg Z-n\pi$. The aim of the following procedure is an iterative updating of the driving frequency $f$ when the material parameters change to remove the phase difference $\Dphi$, i.e., tracking the resonance frequency.

A linear approximation of the phase curve in Eq. \ref{eq:MoDaNE_phase_theory} near resonance can be considered,
\begin{equation}
    \Dphi\of \approx k \left(f-f\res\right),\label{eq:linapprox}
\end{equation}
\noindent which allows estimation of the resonance frequency $f\res$ based on the observed $\Dphi$ and updating the excitation frequency to this estimate (phase feedback). The negative local slope $k$ is given by the resonance frequency and damping (see Eq. \ref{eq:MoDaNE_phase_theory}):
\begin{equation}
    k=-\frac{\pi}{f\res\tanh\alpha L}.\label{eq:k}
\end{equation}
\noindent As material parameters evolve, $k$ also changes, and updating it during the course of the measurement based on estimates of $f\res$ and $\alpha$ from MoDaNE inversion (Eqs. \ref{eq:mod_inv_a} and \ref{eq:mod_inv_p}) is beneficial.

In this paper, we focus on nonlinear variations of the resonance frequency $f\res$ due to the increase in the amplitude of excitation. When a step in excitation amplitude is applied, the material parameters change abruptly and a substantial deviation from the resonance condition $\Dphi=0$ arises. Updating the excitation frequency iteratively based on the observed phase difference to the estimate of the resonance frequency from Eq. \ref{eq:linapprox} will converge to the resonance (i.e. using the phase feedback only), and it is in principle sufficient.  However, anticipating the resonance frequency shift due to the amplitude change and taking it into account together with the measured $\Dphi$ expedites reaching the resonance (amplitude feedforward control). The contribution of this feature to the efficiency of the resonance tracking procedure is demonstrated in Section \ref{sec:rtefficiency}.

In general, the frequency shift caused by the excitation amplitude (voltage $A$) change is unknown (given by an unknown amplitude dependence of frequency variation and time). Here we introduce a local empirical model assuming the frequency shift is proportional to the change in excitation amplitude $\Delta A$:
\begin{equation}
\Df\res=\ell\Delta A.\label{eq:l}
\end{equation}
\noindent The unknown coefficient $\ell$ is to be learned during measurement. In the cases of interest, an amplitude increase produces softening hence $\ell<0$. Note that this linear model does not imply a linear dependence of the resonance frequency on the amplitude as $\ell$ can be interpreted as a local derivative of the amplitude function for the current amplitude.

\subsubsection{Updating rules}

Let us assume that within iteration $i$ the sample is excited with drive frequency $f_i$ at drive amplitude $A_i$ and an output amplitude $a_i$ and phase difference $\Dphi_i$ are measured. Taking into account the two contributions to phase discussed before, we can calculate the frequency $f_{i+1}$ to be used with the following drive amplitude $A_{i+1}$ as
\begin{equation}
    f_{i+1}=f_i-\frac{\Dphi_i}{k_i}+\ell_i\left(A_{i+1}-A_i\right).\label{eq:fup}
\end{equation}
\noindent In the right-hand side of the equation, the second term corrects the current deviation from resonance (phase feedback) and the third term takes into account the following amplitude step (feedforward).

\begin{figure}
	\centering
		\includegraphics[width=0.7\columnwidth]{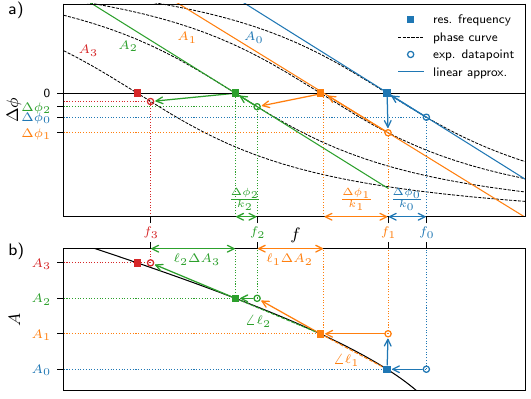}
	\caption{Schematic representation of the initial steps of the procedure. a) Phase vs. frequency at different drive amplitudes (different colors), the circles represent the testing frequency and the corresponding measured phase shift. b) Amplitude dependence of resonance $f\res$ and drive $f$ frequency. The procedure starts at amplitude $A_0$ (blue line) with drive frequency $f_0$ slightly higher (for illustrative purposes) than the resonance frequency $f\res[0]$ (blue square), observing a non-zero $\Dphi_0$ (blue circle). The feedback correction (blue range) is given by the linearization slope $k_0$. $\ell_0$ is unknown, thus feedforward is not applied and $f_1$ is roughly equal to $f\res[0]$. Increasing the amplitude to $A_1>A_0$ (orange line) and testing frequency to $f_1$ gives a significant phase difference $\Dphi_1$ (orange circle) as $f_1$ is far from the resonance frequency $f\res[1]$ (orange square). In this iteration $\ell_1$ can be calculated based on estimates of $f\res[0]$ and $f\res[1]$. Determination of $f_2$ includes both the phase-feedback correction (first orange range) and feedforward adjustment (second orange range), thus measurement at amplitude $A_2$ (green line) and frequency $f_2$ is performed close to the resonance $f\res[2]$ (green square), resulting in a low phase difference $\Dphi_2$ (green circle). The procedure is then repeated for the following amplitudes.}
	\label{fig:proc}
\end{figure}

Phase feedback coefficient  $k_i$ can be determined immediately within the first iteration (or initialized based on linear characterization) by substituting estimates of resonance frequency $f\res[i]$ and damping coefficient $\alpha_i$ (obtained using Eqs. \ref{eq:mod_inv_a} and \ref{eq:mod_inv_p}) into Eq. \ref{eq:k}. The amplitude feedforward coefficient  $\ell_i$ can be initialized and updated only after changing the drive amplitude. The coefficient $\ell$ is defined as a shift of the resonance frequency caused by the change in amplitude divided by the amplitude step. The updating rule can be written as
\begin{equation}
    \ell_i=
    \frac{f\res[i]-f\res[i-1]}{A_i-A_{i-1}} \quad \text{if }A_i\neq A_{i-1}.
    \label{eq:lup}
\end{equation}
The initial steps of the procedure are illustrated in Fig. \ref{fig:proc} and explained step-by-step in the caption.

\subsubsection{Boosting robustness}

In experiments, the updating rules defining $k$ and $\ell$ might be affected by the propagation of errors from the measurements. To limit this, Eqs. \ref{eq:k} and \ref{eq:lup} are replaced with exponential moving averages to smooth the evolution,
\begin{align}
    k_{i} &= -\beta\frac{\pi}{f\res[i]\tanh\alpha_{i} L}+\left(1-\beta\right)k_{i-1}
    \label{eq:kupave}\\
    \ell_{i}&= \beta'\frac{f\res[i]-f\res[i-1]}{A_i-A_{i-1}} + \left(1-\beta'\right)\ell_{i-1}  \;\text{if }A_i\neq A_{i-1}.
    \label{eq:lupave}
\end{align}
\noindent The weight parameters can be tuned to optimize the procedure with respect to measurement noise. In general, $\beta = 1$ is suitable for a scenario with low noise and rapidly changing resonance, low values of $\beta$ are suitable for high noise. In the following, we used $\beta=\beta'=0.5$. An analysis was performed showing that the efficiency of the procedure does not significantly depend on the values chosen.

\subsubsection{Summary of the procedure}

To summarize the resonance tracking procedure, the measurement consists of repeated evaluation of the following algorithm: 
\begin{enumerate}
    \item Acquire the signals, calculate $a_i$ and $\Dphi_i$ and estimate $f\res[i]$ and $\alpha_i$ using MoDaNE inversion (Eqs. \ref{eq:mod_inv_a} and \ref{eq:mod_inv_p}).
    \item Update the phase slope $k_i$ using Eq. \ref{eq:kupave}.
    \item Update the amplitude coefficient $\ell_i$ using Eq. \ref{eq:lupave}.
    \item Update the excitation frequency $f_{i+1}$ using Eq. \ref{eq:fup}.
    \item Set the generator to the excitation frequency $f_{i+1}$ and the amplitude $A_{i+1}$.
    \item Wait until the transient phase subsides and repeat.
\end{enumerate}

\section{Demonstration}\label{sec:demo}

The proposed method is demonstrated through the testing of a sandstone sample and the resonance tracking NRUS. We show here that the procedure keeps a constant phase shift during the experiment even when the material properties are evolving in time due to slow dynamic effects. 

\subsection{Experimental setup}\label{sec:expsetup}

\begin{figure}
	\centering
		\includegraphics[width=0.7\columnwidth]{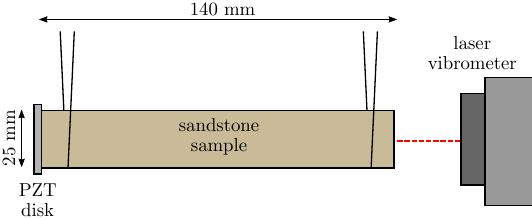}
	\caption{Experimental setup.}
	\label{fig:expsetup}
\end{figure}

The experimental data are obtained using the experimental setup illustrated in Fig. \ref{fig:expsetup}. A prismatic sandstone sample ($140 \times 25 \times 25 \:\unit{\milli\meter}$, see \cite{kober_role_2025} for material parameters) is suspended horizontally in order to approximate free boundary conditions. 

The sample is excited using a PZT disk (APC $20 \times 2 \:\unit{\milli\meter}$ ring) glued to one of the bases and driven using a Tabor 9400 power amplifier (up to \qty{150}{\volt}). We tracked the resonance frequency of the first longitudinal mode at approx. \qty{8.5}{\kilo\hertz}. The amplifier monitor output (high amplitude output divided by 100) is used as the $y\tx$ signal. A Polytec OFV-505 laser vibrometer with OFV-5000 controller is used to measure the surface velocity of the free end, the response signal $y\rx \sim \dot{u}\left(L\right)$. The velocity is converted into strain in the center of the sample using the linear approximation $\varepsilon=\dot{u}\left(L\right) / v_\text{L}$, where $v_\text{L}$ is the linear longitudinal wave velocity.

The transducer (PZT disk) and vibrometer are connected to a Keysight / Signadyne M3300A arbitrary waveform generator and oscilloscope, respectively. The signal generator allows continuous adjustment of frequency and amplitude without phase discontinuities, ensuring uninterrupted excitation. The signals $y\rx$ and $y\tx$ are recorded in \qty{5}{\milli\second} pieces with a sampling rate of \qty[per-mode = symbol]{5}{\mega\sample\per\second}. When an amplitude or a frequency change is performed, these signals are recorded after a delay allowing the system to reach standing wave conditions. The duration is defined from the damping properties of the material, so that \qty{99}{\percent} of the transient amplitude effect subsides: $t_{\qty{99}{\percent}}=-\log \left(0.01\right)/\left(\alpha v_\text{L}\right)$. For the tested material $t_{\qty{99}{\percent}} = \qty{10}{\milli\second}$. Such timing was used in all of the following, whenever monochromatic excitation was applied, unless explicitly stated otherwise.

\subsection{Linear characterization}\label{sec:calibration}

The linear characterization (calibration) is performed only once before the actual measurement starts, by measuring the resonance on a low amplitude, providing estimates of the parameters used in the MoDaNE solution. It allows compensation for effects of the experimental setup which make the experimentally observed data differ from the theory due to a frequency dependent amplitude and phase response (e.g. the gain of the amplifier or the transducers response, delays in the signals transmission and reception, etc.). Assuming experiments are performed in a small frequency range, corrections (linear in frequency), which are multiplicative for amplitude and additive for phase, have been introduced. It follows:
\begin{align}
    a\ex\of
    &=\frac{q_0+q_1 f}{\sqrt{\cosh^2\left(\alpha L\right)-\cos^2 \left(\pi f/f\res\right)}}, \label{eq:MoDaNE_a}\\
    \varphi\ex\of
    &=p_0+p_1 f-\arctan\left[\frac{\tan\left(\pi f/f\res\right)}{\tanh \left(\alpha L\right)}\right],\label{eq:MoDaNE_phase}
\end{align}
\noindent where $q_0+q_1 f$ represents the displacement amplitude in $x=0$ and $p_0+p_1 f$ accounts for the phase shift and the delay in time (which is linear in frequency). If the normalized formulation of MoDaNE is used (applying it to the transfer function instead of displacement), the parameters do not depend on the excitation amplitude.

The resonance peak is sampled using monochromatic excitation at very low amplitude (\qty{1}{\volt}) covering a bandwidth of \qty{1.6}{\kilo\hertz} with steps of \qty{8}{\hertz}. The resonance curve (Fig. \ref{fig:res}) is then used to estimate (corrected) MoDaNE parameters in Eqs. \ref{eq:MoDaNE_a} and \ref{eq:MoDaNE_phase}.

The results from calibration are then applied for each amplitude of excitation. Having measured the response $Z$ on excitation frequency $f$, the theoretical amplitude and phase are
\begin{equation}
a=\frac{\lvert Z\rvert}{q_0+q_1 f}\; \text{and}\;\varphi=\arg Z-p_0-p_1 f+\pi n,\label{eq:calibinversion}
\end{equation}
\noindent These values can be analytically inverted for the estimation of the resonance frequency and damping using MoDaNE inversion (Eqs. \ref{eq:mod_inv_a} and \ref{eq:mod_inv_p}).  We also redefine $\Dphi\equiv\arg Z-p_0-p_1 f$ for the purpose of the resonance tracking procedure.

\subsection{Resonance tracking implementation}\label{sec:demo-implementation}

\begin{figure*}
	\centering
		\includegraphics[width=\textwidth]{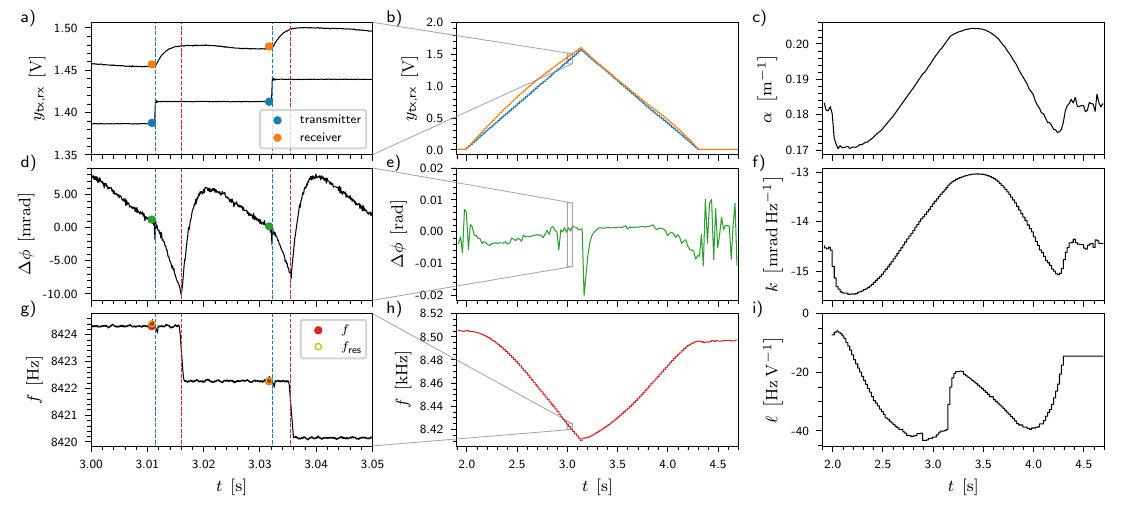}
	\caption{Measurement protocol and resonance tracking. Subplots in second and third column contain experimental data from the resonance tracking procedure during the entire course of experiment. First column shows a zoomed-in short window that includes both data from the resonance tracking (points) and from the continuous acquisition (lines). ab) Excitation (blue) and measured (orange) amplitudes. The real transmitter voltage is $100\times$ higher. de) Phase difference from resonance tracking (green) and from continuous acquisition (black). gh) Excitation frequency (dots and red line) and estimate using MoDaNE inversion (circle). c) Damping estimate using MoDaNE inversion. f) Phase feedback coefficient. i) Amplitude feedforward coefficient.}
	\label{fig:protocol}
\end{figure*}

The nonlinear response of the sample is measured using a sequence of uniformly spaced excitation amplitudes (monochromatic waves) that includes loading (increasing) followed by unloading (decreasing). In addition, a continuous acquisition at \qty[per-mode = symbol]{1}{\mega\sample\per\second} of the input and response signals is used to evaluate the amplitude and phase shift during the whole measurement process. This supplementary data provide information about the response of the system to the resonance tracking procedure.

The course of the experiment, including all resonance tracking-related quantities, is shown in Fig. \ref{fig:protocol}, where the full temporal evolution is shown in the central column and a zoom in a short time window is reported in the left column using continuously recorded data. Results from continuous acquisition are shown as solid lines, where results from the discrete acquisition are reported as symbols. The excitation amplitude $A$ starts at a low amplitude (\qty{0.5}{\volt}) and it grows up to \qty{150}{\volt} with a step of approx. \qty{2.5}{\volt}. After reaching its maximum, $A$ decreases to the initial amplitude. For each amplitude, the nominal source frequency $f$ is defined applying the resonance tracking procedure. 

Within the \qty{15}{\milli\second} duration of the excitation at a given amplitude and frequency (see the first column of Fig. \ref{fig:protocol} to appreciate details), we can observe the time instances at which the amplitude is changed (vertical blue dashed lines). Slightly later, the frequency at the generator is changed (vertical dashed red lines), as a result of the phase tracking calculations, and kept constant. The slight delay between the amplitude and the frequency change is due to signal processing that is performed after changing the amplitude. As soon as the drive amplitude increases, the resonance frequency starts dropping (conditioning time regulates the time needed to achieve the new value), thus $f>f\res$, and a negative phase appears. Later, $f$ is updated through the resonance tracking protocol and the behavior of the phase follows the adjustment of $f$ to $f\res$, mixed up with some effects due to the transient condition. At the end of each iteration, the strain reaches a steady state, and the phase difference falls to a proximity of zero. This agrees with the results of the resonance tracking given by the circles just before the amplitude change. The behavior is highly repeatable in the course of the experiment.

Overall, each amplitude step corresponds to one iteration of the procedure and yields one datapoint (strain, phase, testing frequency), from which material properties can be calculated using MoDaNE equations (resonance frequency $f\res$ and damping $\alpha$) and the parameters for resonance tracking can be updated ($k$ and $\ell$). Their temporal evolutions are reported in the second and third column of Fig. \ref{fig:protocol}. The amplitude coefficient $\ell$ (Fig. \ref{fig:protocol}i) is initially undefined and is learned when the amplitude is changed. The resonance tracking algorithm maintains the phase difference $\Dphi$ close to zero (Fig. \ref{fig:protocol}e) by decreasing the excitation frequency (Fig. \ref{fig:protocol}h), as the material is softening during the loading branch and vice versa during unloading. The estimates of the resonance frequency (not shown) are almost superimposed on the excitation frequency. The estimates of material parameters are used to update the phase curve slope $k$ (Fig. \ref{fig:protocol}f), which is correlated with $\alpha$ (Fig. \ref{fig:protocol}c) as the slope is affected particularly by damping (Eq. \ref{eq:k}). The amplitude coefficient $\ell$ (Fig. \ref{fig:protocol}i), defined using Eq. \ref{eq:l}, approximates the derivative of the resonance frequency with respect to excitation amplitude, thus it can be noticed to decrease as long as the dependence (see results shown later in Fig. \ref{fig:ampfun}a) gets steeper, i.e. where the derivative of the resonance frequency with respect to strain increases.

\subsection{Efficiency of the resonance tracking procedure}\label{sec:rtefficiency}

\begin{figure}
	\centering
		\includegraphics[width=0.7\columnwidth]{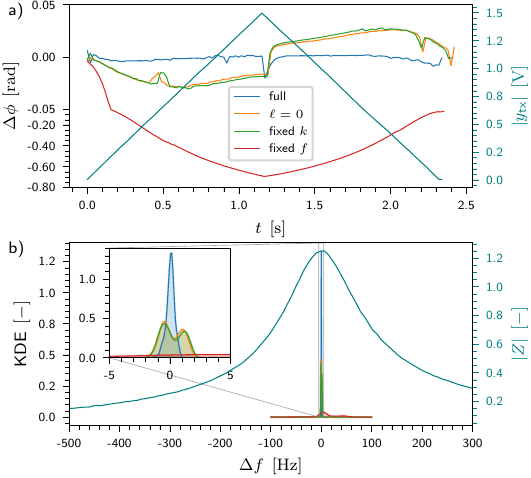}
	\caption{Phase difference (a) and distribution of excitation frequency deviations from the resonance obtained as kernel density estimation (b) during a measurement with increasing and decreasing amplitudes (cyan) performed without the resonance tracking (red), with a fixed phase slope $k$ and no amplitude feedforward (green), with adaptive $k$ and no amplitude feedforward (orange) and using full resonance tracking with the amplitude feedforward (blue).}
	\label{fig:phasealgs}
\end{figure}

To demonstrate the contribution of individual components of the resonance tracking algorithm, the very same experiment was repeated but disabling one by one the features used for resonance tracking. As illustrated in Fig. \ref{fig:phasealgs}a, measuring at a fixed frequency (i.e. not applying any resonance tracking) results in an increase in the phase difference $\Dphi$ to more than \qty{0.7}{\radian} as the excitation amplitude increases, which is expected as the resonance frequency drops. In the absence of amplitude feedforward control ($\ell$ fixed to zero), phase differences approximate zero, with slightly superior outcomes observed for adaptive phase slope $k$ (i.e. when a flattening of the phase curve due to increasing damping is taken into account). However, the phase difference is consistently negative when the amplitude is increasing (the resonance frequency is overestimated) and positive during unloading (the resonance frequency is underestimated). Resonance tracking without feedforward control is always one step behind. The issue is resolved through the implementation of amplitude feedforward (full resonance tracking): initially the phase difference follows roughly the same curve as the previous cases but when the coefficient $\ell$ is learned during the process, it stabilizes near zero, exhibiting no offset between the loading and unloading branches. Adjusting the phase curve slope and predicting the effect of the amplitude steps indeed both contribute to the efficiency of the resonance tracking.

Fig. \ref{fig:phasealgs}b shows the distribution (kernel density estimation) of the deviation of the excitation frequency from the resonance, $\Delta f=f-f\res$ for the same data as in Fig. \ref{fig:phasealgs}a, illustrating the relative position on the resonance peak.

\section{Applications}\label{sec:applications}

Typical applications of the resonance tracking based NRUS procedure are given here to demonstrate the usefulness and advantages of the proposed procedure, in particular with respect to conventional NRUS methods.

\subsection{Resonance tracking NRUS}

The experiment described in the previous section allows to analyse the strain dependence of resonance frequency and damping. The results of NRUS performed using the resonance tracking approach are compared with two other NRUS methods. The conventional NRUS measurement \cite{van_den_abeele_micro-damage_2001}, i.e., using a sequence of monochromatic excitations, is performed with the same measurement timing as resonance tracking. The chirp NRUS \cite{maier_noncontact_2018} uses a linear chirp signal of a duration of \qty{1}{\second} tapered with a Tukey window. Both methods use the same frequency bandwidth and amplitude sequence.  The frequency bandwidth is \qty{800}{\hertz} (with a step of \qty{5}{\hertz} for conventional NRUS sweep). The frequency sweeps are repeated at amplitudes starting from \qty{0.5}{\volt}, going up to \qty{150}{\volt} in 60 equal steps (loading) and then decreasing through the same values (unloading). The experimental data are shown in Fig. \ref{fig:nrus}. 

The measurements were taken successively ensuring constant environmental and boundary conditions, though not necessarily initiated in a relaxed state and/or in the same conditioning state. The measurement durations differ significantly: resonance tracking is complete within \qty{2.5}{\second}, chirp measurement takes approximately \qty{2}{\minute}, and conventional NRUS was measured in \qty{5}{\minute}.

The resonance frequency and damping are consistently estimated using the MoDaNE inversions for all of the three approaches. In the case of sine NRUS, the signal measured with the highest amplitude is used for the estimation of the signal amplitude and phase, i.e. the datapoint with a frequency closest to the resonance. The chirp measurements are processed using a discrete Fourier transform with a resolution of \qty{2}{\hertz} and the amplitude and phase values corresponding to the highest amplitude are used in the MoDaNE inversion.

\begin{figure}
	\centering
		\includegraphics[width=0.7\columnwidth]{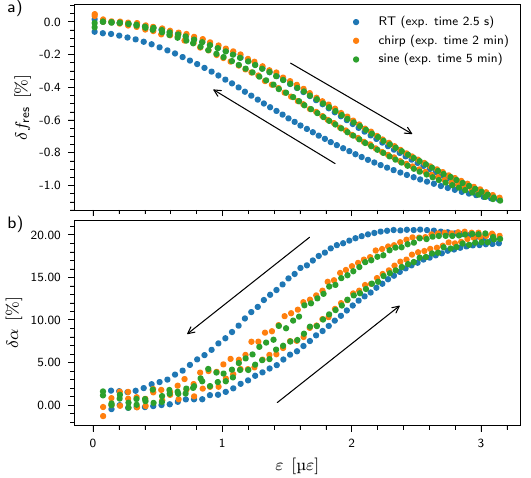}
	\caption{Comparison of NRUS results measured using resonance tracking, monochromatic excitations and chirps. a) Relative resonance frequency variation (defined as $\delta f\res=\left(f\res-f\res[0]\right)/f\res[0]$ where $f\res[0]$ is the resonance frequency at the lowest amplitude), b) damping coefficient variation (defined as $\delta \alpha=\left(\alpha-\alpha_0\right)/\alpha_0$ where $\alpha_0$ is the damping at the lowest amplitude).}
	\label{fig:ampfun}
\end{figure}

The strain dependences of both relative resonance frequency shift $\delta f\res$ and relative change in damping $\delta \alpha$ obtained using the three NRUS approaches are reported in Fig. \ref{fig:ampfun}. The curves generally agree with results reported in the literature for the loading branch: an initial quadratic behavior is followed by an approximately linear dependence. We observe a good agreement between the methods both in shape and values. On the contrary, the unloading branch is different in the three cases, where the resonance tracking shows larger loop area. The effect is likely due to a different conditioning and relaxation. Since the duration of measurements in resonance tracking is much shorter than in the other cases, it is expected slightly less conditioning during loading (explaining the small differences in the loading branch) and significantly less relaxation during unloading (leading to the significantly higher values of velocity and damping variations during unloading in resonance tracking). Notably, the damping still slightly grows even though the strain decreases at the beginning of the unloading. Thanks to the very short acquisition time, resonance tracking allows to explore behaviors which cannot be tracked with standard NRUS methods.

\subsection{Effects of conditioning duration}

\begin{figure}
	\centering
		\includegraphics[width=0.7\columnwidth]{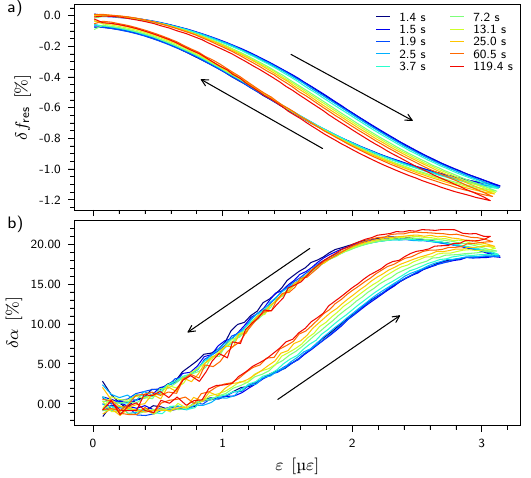}
	\caption{Results obtained using resonance tracking in measurements with various durations. a) Relative resonance frequency variation, b) damping variation. Color scale represents the total duration of the measurement.}
	\label{fig:varpause}
\end{figure}

The acquisition time can be controlled in resonance tracking and kept minimal, thus making it an optimal technique to study the role of the duration of the excitation for what concerns conditioning (and thus also supporting the discussion of the previous figures). To this purpose, the same resonance tracking based NRUS measurement was performed using various rates. The duration of the excitation for each amplitude was varied from \qty{6}{\milli\second} to \qty{1}{\second} resulting in measurements performed within $T_\text{exp}=\qty{1.4}{\second}$ to \qty{2}{\minute}, with the fastest measurement not performed in the stationary state. The signals used for the analysis always capture the last \qty{5}{\milli\second} segment of excitation (i.e. just before switching to the following amplitude). The resulting strain dependences of relative resonance frequency shift $\delta f\res$ and relative change in damping $\delta \alpha$ are shown in Fig. \ref{fig:varpause}. 

\begin{figure}
	\centering
		\includegraphics[width=0.7\columnwidth]{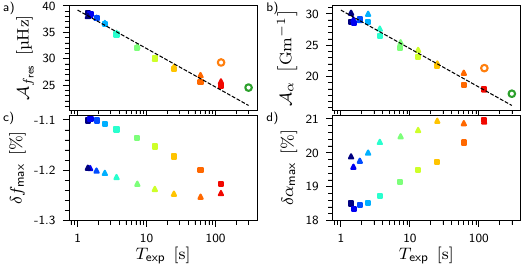}
	\caption{Results obtained using resonance tracking in measurements with various durations. ab) Areas $\mathcal{A}_{f\res}$ and $\mathcal{A}_\alpha$ enclosed by $f\res\oeps$ and $\alpha\oeps$ curves vs. measurement duration $T_\text{exp}$, cd) resonance frequency and damping variations measured at the highest excitation amplitude vs. $T_\text{exp}$. Squares correspond to measurements obtained from experiments repeated one after the other increasing the experiment duration (data shown in Fig. \ref{fig:varpause}) while triangles from experiments decreasing durations. Green and orange circles show, for reference, the values obtained with conventional and chirp-based NRUS.}
	\label{fig:varpause2}
\end{figure}

The nonlinearity and loading--unloading asymmetry of each curve are quantified by the relative resonance and damping changes at the strongest excitation amplitude ($\delta f_\text{max}$ and $\delta\alpha_\text{max}$) and the respective curve opening (area enclosed by the loading and unloading curve, $\mathcal{A}_{f\res}$ and $\mathcal{A}_\alpha$). These quantities  are shown in Fig. \ref{fig:varpause2} as functions of measurement time. Here, data are shown for two modes of the experiment: first, the duration is increased (circles), while in the second it is decreased (triangles). Note the good agreement between the two experiments. 

We observe that the nonlinear effect (decrease in resonance frequency and increase in damping) during loading becomes stronger with increasing experiment duration (blue to red) and the opposite during unloading (Fig. \ref{fig:varpause}). A slower measurement produces more conditioning during loading, but also more relaxation during unloading, therefore the curves are less open and area decreases. The values of $\delta f_\text{max}$ and $\delta\alpha_\text{max}$ show an aprroximate 10 \% variance in the range of $T_\text{exp}$ tested. Similar sensitivity to measurement duration would be observed for nonlinear parameters derived from the slopes of the $\delta f\res$ and  $\delta \alpha$ curves.

We finally remark that the results of slow measurements do not match those obtained using conventional approaches to NRUS, despite sharing the same duration, as the strain amplitude varies during frequency sweep and the average conditioning amplitude is lower. Furthermore, the strain spatial profiles during measurement vary with frequency in conventional NRUS, while this is not occurring in resonance tracking NRUS implementation.

\subsection{Monitoring of conditioning and relaxation}\label{sec:relaxation}

The resonance tracking can be applied to a measurement involving conditioning using excitation at constant amplitude and relaxation process. In this case, the temporal evolution of the material parameters is analysed, i.e. slow dynamics is studied.

\begin{figure*}
	\centering
		\includegraphics[width=\textwidth]{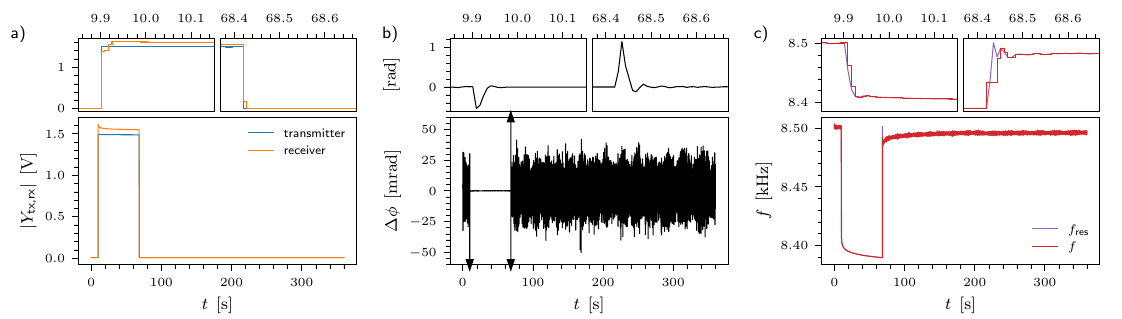}
	\caption{Conditioning and relaxation experiment. a) Excitation (blue) and measured (orange) amplitude vs. time. The real transmitter voltage is $100\times$ higher. b) Phase difference vs. time. c) Excitation frequency (red) and resonance frequency estimated using MoDaNE inversion (violet). Zoom of beginning of conditioning and relaxation are shown in the upper row.}
	\label{fig:cndrlx}
\end{figure*}

As conventionally done, to estimate the material parameters, we measure the preconditioning and relaxation phases exciting with a monochromatic wave at very low drive amplitude (\qty{0.5}{\volt}). Between them, the sample is excited at large amplitude.  The drive amplitude protocol is shown in Fig. \ref{fig:cndrlx}a (blue). During conditioning and relaxation phases, the resonance frequency evolves in time, even though the  drive amplitude is constant, thus requiring resonance tracking. The drive/estimated resonance frequencies are shown versus time in Fig. \ref{fig:cndrlx}c and match well.

Contrary to the previous experiments, the \qty{5}{\milli\second} signals were recorded almost continuously, i.e. the \qty{10}{\milli\second} delay between successive acquisitions was omitted. This increased the time resolution of both conditioning and relaxation processes. The delays between acquisitions were given only by the equipment latency and processing time. Consequently, resonance tracking keeps the sample consistently near standing wave conditions, except immediately after changing the drive amplitude. We expect significant effects particularly at the very beginning of the relaxation measurement, when the received signal consists predominantly of the exponential decay of the preceding high-amplitude excitation (ring-down). Thus, the amplitude and phase are not measured reliably and should not be used in the analysis. We adopt an inhibition mechanism that makes the excitation frequency and resonance tracking parameters fixed for a given number of iterations (2 in our case), i.e. for the first approx. \qty{10}{\milli\second} following a drop in amplitude. This mechanism mimics the pause applied in the amplitude-dependent NRUS protocol, allowing high temporal resolution while ensuring tracking stability. Also note that, due to the constant amplitude excitation, resonance tracking is performed mostly using phase feedback only, while amplitude feedforward contributes only at the onset of relaxation.

The experimental protocol is illustrated in Fig. \ref{fig:cndrlx}: after a short (\qty{10}{\second}) preconditioning phase, the sample is excited at high amplitude (results for $A=\qty{150}{\volt}$ are shown here) for \qty{60}{\second} (conditioning phase) and the following relaxation process is measured for \qty{1}{\hour}. The phase difference (Fig. \ref{fig:cndrlx}b) remains almost zero during the whole experiment, except for a short time when the amplitude is changed, see the upper row. The high variances during preconditioning and relaxation are caused by low signal-to-noise ratios of the low excitation amplitude signal. This is a further demonstration of the efficiency of the resonance tracking procedure dealing with an abrupt change in the material parameters.

During conditioning ring-up (when the excitation amplitude is increased), the resonance frequency gradually decreases as the strain goes up. The amplitude feedforward is not applied here (assuming the coefficient is still unknown). Consequently, the drive frequency update is "delayed" with respect to the resonance frequency (Fig. \ref{fig:cndrlx}c) causing a peak in the phase difference that is rapidly corrected in time using the phase feedback.

The ring-down (very initial steps of relaxation) case is different. The amplitude feedforward is applied here (note the simultaneous increase in drive frequency and drop in amplitude), but the phase feedback cannot be used as the experimental data are contaminated by the ring-down conditioning. As the drive frequency is not updated for a few iterations, the initial spike of the phase difference (Fig. \ref{fig:cndrlx}b) and of the estimated resonance frequency (Fig. \ref{fig:cndrlx}c) were expected and meaningless.

\begin{figure}
	\centering
		\includegraphics[width=0.7\columnwidth]{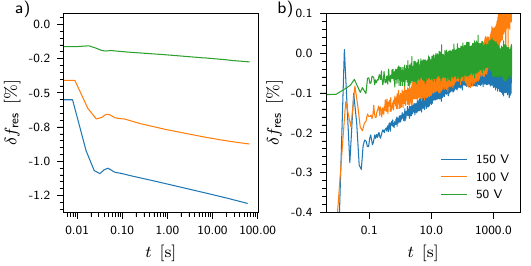}
	\caption{Relative resonance frequency variations during a) conditioning, b) relaxation. Datasets refer to different conditioning amplitudes.}
	\label{fig:cndrlxlog}
\end{figure}

The measurement described above was performed using several conditioning amplitudes. The resonance frequency during both conditioning and relaxation processes (Fig. \ref{fig:cndrlxlog}) exhibits a linear evolution in log-time as described in the literature.

\section{Discussion}\label{sec:discussion}

\subsection{Consideration on the resonance tracking procedure}

The proposed approach, implementing a discrete-time resonance tracking algorithm, is designed to maintain a dynamical system vibrating with its resonance mode during an experiment involving varying material parameters. Rather than repeatedly sampling a resonance curve, the excitation frequency is iteratively updated so that a prescribed phase condition between excitation and response is satisfied.

For each iteration, the system is driven at a selected frequency and amplitude for a predefined duration, allowing the transient wave to decay. The response is then evaluated, and the phase difference between excitation and response is computed. Based on this phase deviation from the resonance condition, the excitation frequency for the next iteration is updated according to a model of the local frequency--phase relation, linearized to improve robustness to measurement noise and stability. An analytical model of linear monochromatic standing waves  (MoDaNE) is used to adapt the phase--frequency slope based on damping estimate, which is shown to contribute to the efficiency of the procedure even in the nonlinear regime. In this manner, the algorithm performs a feedback correction that converges toward and subsequently tracks the evolving resonance frequency.

A feedforward control is implemented in the procedure to predict the changes in the resonance frequency as a response to the varying excitation amplitude due to fast nonlinear effects. However, the material parameters evolve also due to other effects: slow dynamics and possible sample temperature increase caused by energy dissipation (as observed in \cite{zeman_distribution_2025}). As the amplitude sequence is constructed using uniform steps and the timing of the iterations is uniform as well, the feedforward control implicitly accounts for these slow effects too, extrapolating the resonance frequency evolution in time. This behaviour is beneficial for the efficiency of the resonance tracking; however, interpreting the coefficient of the feedforward term as a derivative of the amplitude dependence is not fully correct. This effect might be partially responsible for the asymmetry in the coefficient $\ell$ between loading and unloading (Fig. \ref{fig:protocol}i).

The proposed algorithm shares conceptual similarities with phase-locked loop (PLL) techniques \cite{denis_identification_2018} widely used for resonance tracking in nonlinear dynamical systems. However, the present implementation differs in several key aspects. First, the procedure operates in discrete time, ensuring that the frequency update is based on quasi-stationary conditions. Second, the update rule incorporates a local linearized model with adaptive gain and optional feedforward terms associated with controlled experimental parameters. The method is therefore not formulated as a continuous feedback controller but as an iterative measurement protocol tailored to systems with evolving material properties.

Furthermore, the tracking framework is inherently independent of the specific mode shape or order; as long as the selected resonance peak is sufficiently isolated from adjacent modes to prevent phase interference.

\subsection{Considerations about MoDaNE}\label{sec:discussion-modane}

Although the MoDaNE formulation is used throughout this work as a convenient reference model, the proposed resonance tracking method does not rely on the validity of the MoDaNE solution in nonlinear or heterogeneous materials. In particular, the definition of resonance employed here is based solely on the experimentally observed phase response and does not require any assumption of linearity or spatial homogeneity.

As demonstrated experimentally (Fig. \ref{fig:nrus}), the resonance frequency defined by the zero crossing of the phase remains consistent with the conventional amplitude-based definition used in conventional NRUS, even at elevated excitation amplitudes where nonlinear effects and slow dynamics are present. The MoDaNE model is therefore used primarily as a supporting tool rather than as a fundamental assumption underlying the method.

Within this framework, the MoDaNE equations serve two practical purposes. First, they provide a simple parametrization for correcting linear frequency-dependent distortions of amplitude and phase introduced by the experimental setup. In particular, the linear phase correction inferred from the model was found to accurately match the nominal delay of the vibrometer, confirming its instrumental origin.

Second, the model offers an estimate of the local slope of the phase--frequency relation near resonance, which is used to update the proportionality coefficient in the frequency update step. Importantly, the resonance tracking procedure does not require an accurate estimate of this slope to function correctly; even approximate values lead to stable tracking. The adaptive update merely improves convergence speed and robustness, as demonstrated in Section \ref{sec:rtefficiency}.

In the MoDaNE formulation, the attenuation coefficient represents purely dissipative losses in a linear, homogeneous medium. In the present experiments, however, the estimated attenuation is derived from the amplitude of the fundamental harmonic and therefore reflects all mechanisms that reduce energy at the excitation frequency. As a consequence, the measured attenuation should be understood as an effective attenuation of the fundamental response, incorporating not only intrinsic dissipation but also energy transfer to higher harmonics and other nonlinear processes. While this quantity does not correspond to a purely dissipative material parameter, it remains a meaningful and reproducible indicator of the system's response and its evolution with excitation amplitude. Also, it is merely an average as an inhomogeneity is induced in the sample by slow dynamics \cite{kober_role_2025}.

\subsection{Benefits of resonance tracking based measurements}

The approach described here presents two main advantages. Performing a measurement exciting the sample using the resonance frequency only for each excitation amplitude ensures that the spatial distribution of strain remains associated with a single dominant mode (first longitudinal mode in the case of this paper). Second, reducing the measurement time to a minimum allows improving time resolution and reducing the influence of cumulative conditioning associated with long acquisition times.

Probing slightly different modal mixtures at different drive amplitudes when the excitation frequency deviates from resonance, leads to variations in the spatial strain distribution, thus comparing results in different probing conditions. Although the modal distortions are typically small around the resonances, their influence may become non-negligible, particularly in materials with slow dynamic effects, where induced heterogeneity would accumulate and affect following probings. Besides issues related to modal distortions, resonance tracking ensures that when drive amplitude increases/decreases the sample is constantly probed at increasing/decreasing strain. This cannot be guaranteed when the probing frequency $f$ does not always match the resonance. Oscillations of the probing frequency around the resonance cause decrease/increase in strain even though the drive amplitude is not varied. In materials with slow dynamics that means switching from conditioning to relaxation within the duration of a single experiment. For instance, sampling the whole peaks at a given drive amplitude involves huge variations in the strain applied and also periods during which strain increases/decreases (before/after $f=f\res$). Avoiding such effects is crucial for reliable/interpretable measurements.

Avoiding probing of the whole frequency range for each amplitude allows to decrease the time required to complete the measurement by orders of magnitude without losing information about the resonance frequency or damping. Recall that the out-of-resonance datapoints are not used even in conventional measurements, except eventually for the estimation of Q factor from the peak width. The reduction in experimental time is advantageous from several points of view:
\begin{itemize}
    \item Recent study \cite{kober_role_2025} indicates that prolonged probing enhances the contribution of slow dynamics, thereby increasing protocol dependence. Also the impact of variations of material parameters due to uncontrolled environmental conditions or increasing temperature caused by energy dissipation in the sample increases with the duration of experiments. The influence of the probing time on the results is confirmed from our data (Fig. \ref{fig:varpause}): as the experiment duration increases, the nonlinear effect during loading becomes stronger due to higher contribution of the slow dynamics effects caused by longer excitation, while during the unloading branch the prolonged measurement on decreasing amplitudes allows the material to relax more, in agreement with observations reported elsewhere using the Scaling Subtraction Method \cite{scalerandi_role_2020}. Consequently, the slowly performed measurements provide a nonlinear response more similar to those obtained using sine and chirp NRUS measurements that are performed with a similar experiment time.
    \item The interplay between conditioning (very slow and cumulative) and nonlinearity (fast) effects can be resolved only with a very fast protocol that does not allow conditioning effects to build up. Our data show the largest apparent damping and frequency shifts during loading observed in conventional NRUS (see Fig. \ref{fig:ampfun}), because of the longer duration of the experiment, as confirmed by the cumulative conditioning observed with resonance tracking for damping at the beginning of the unloading phase ($\alpha$ still increases while driving amplitude already decreases), also observed in other fast measurements using DAET \cite{kober_role_2025}. For a correct interpretation of the data it is thus crucial to track whether or not the material reached an equilibrium at each loading step. Likely, different protocols probe slightly different effective states of a nonlinear material, having in mind that overlapping of phases of dominant relaxation or dominant conditioning, as discussed in the previous paragraphs, does not help in assessing the actual state of the material.
    \item Tracking the evolution of the material parameters close to a significant jump in amplitude requires a temporal resolution in the measurement sufficiently high and at the same time avoids contamination due to temporal variations in strain profile or other features which might occur in the same temporal interval. Resonance tracking seems to be effective also in this direction (see results in Fig. \ref{fig:cndrlx}) even though the application here still needs to be better evaluated.
\end{itemize}

\subsection{Other applications}\label{sec:discussion-otherapp}

Although the applications presented in this work focus on Nonlinear Resonant Ultrasound Spectroscopy, the proposed resonance tracking framework is not limited to amplitude-dependent measurements. The method is, in principle, applicable to any experiment in which an externally controlled parameter modifies the resonance frequency, e.g., temperature variations. In such cases, the controlling quantity would not be the excitation amplitude but the prescribed sample temperature, while the resonance tracking algorithm would still adjust the excitation frequency to maintain the prescribed phase condition.

Beyond externally controlled parameters, the feedforward component may also incorporate time as an explicit variable if the resonance frequency evolves slowly due to conditioning, aging, or other gradual processes. In such cases, the frequency update extrapolates the recent trend, effectively predicting the expected resonance shift between successive iterations. This approach is particularly advantageous when the resonance frequency drifts monotonically and on time scales longer than the iteration loop. However, the usefulness of time-based feedforward control depends on the signal-to-noise ratio: only frequency changes that exceed the noise level between iterations can be reliably extrapolated. 

The proposed approach represents in our opinion a general strategy for adaptive resonance measurements rather than a method restricted to nonlinear amplitude-dependent spectroscopy.

\section{Conclusions}

We have presented a model-assisted discrete-time resonance tracking framework that maintains a resonant system at its instantaneous resonance condition using phase-based frequency updates. Unlike conventional sweep-based approaches, the method does not require the acquisition of complete resonance curves and therefore reduces the sensitivity to measurement duration, transient buildup, and protocol-dependent cumulative effects.

The algorithm includes two key elements: a phase-based definition of resonance with adaptive linear frequency update rule, and an optional feedforward term that improves efficiency when externally controlled parameters systematically shift the resonance frequency.

Application to Nonlinear Resonant Ultrasound Spectroscopy demonstrates that resonance tracking yields consistent resonance frequencies and damping estimates while limiting distortions caused by slow dynamic material evolution and mode reshaping. The measured parameters still depend on the conditioning state; resonance tracking merely gives improved control over that state by shortening, controlling and defining the acquisition protocol. In conditioning--relaxation experiments, the approach enables continuous monitoring of resonance frequency and damping even during low-amplitude probing, providing a unified measurement strategy.

Although motivated by NRUS, the method is general and applicable to a broad class of resonant systems with evolving parameters. By combining model-informed updates with explicit control of experimental timing, the proposed framework offers a robust alternative to sweep-based or purely analog tracking techniques.

\section*{Acknowledgements}
J.~K. and R.~Z. acknowledge the financial support provided by the Ministry of Education, Youth, and Sports of the Czech Republic via the project No. CZ.02.01.01/00/23\_020/0008501 (METEX), co-funded by the European Union. J.~K. and R.~Z. are funded by the institutional support through the grant RVO: 61388998.

\section*{Data availability}
The data reported in this study are openly available in Zenodo at doi.org/10.5281/zenodo.19604874 \cite{zeman_data_2026}.

\bibliographystyle{elsarticle-num-names}
\bibliography{zotero}

\end{document}